\documentclass[conference, a4paper]{IEEEtran}
\IEEEoverridecommandlockouts
\usepackage{cite}
\usepackage{amsmath,amssymb,amsfonts}
\usepackage{algorithmic}
\usepackage{graphicx}
\usepackage{xcolor}
\usepackage{gensymb}
\usepackage[font=small,labelfont=bf]{caption}
\usepackage{subcaption}
\usepackage[left=1.57cm,right=1.57cm,top=0.95cm,bottom=2.54cm]{geometry}

\def\BibTeX{{\rm B\kern-.05em{\sc i\kern-.025em b}\kern-.08em
    T\kern-.1667em\lower.7ex\hbox{E}\kern-.125emX}}
\begin{document}

\title{Simulation Study on Super-Resolution for Coded Aperture Gamma Imaging\\
{\footnotesize}
\thanks{The authors gratefully acknowledge the data storage service SDS@hd supported by the Ministry of Science, Research and the Arts Baden-W{\"u}rttemberg (MWK) and the German Research Foundation (DFG) through grant INST 35/1314-1 FUGG and INST 35/1503-1 FUGG. This paper was funded by the Zentrales Innovationsprogramm Mittelstand (ZIM) under grant KK5044701BS0.}
}

\author{
\IEEEauthorblockN{1\textsuperscript{st} Tobias Mei{\ss}ner}
\IEEEauthorblockA{\textit{Mannheim Institute for Intelligent Systems in Medicine}\\
Heidelberg University\\
Mannheim, Germany \\
tobias.meissner@medma.uni-heidelberg.de} \\
\IEEEauthorblockN{3\textsuperscript{rd} J{\"u}rgen Hesser}
\IEEEauthorblockA{\textit{Mannheim Institute for Intelligent Systems in Medicine} \\
\textit{Interdisciplinary Center for Scientific Computing (IWR)} \\
\textit{Central Institute for Computer Engineering (ZITI)} \\
\textit{CZS Heidelberg Center for Model-Based AI} \\
Heidelberg University\\
Mannheim, Germany \\
juergen.hesser@medma.uni-heidelberg.de}
\and
\IEEEauthorblockN{2\textsuperscript{nd} Werner Nahm}
\IEEEauthorblockA{\textit{Institute of Biomedical Engineering} \\
Karlsruhe Institute of Technology (KIT)\\
Karlsruhe, Germany \\
werner.nahm@kit.edu} \\
\IEEEauthorblockN{4\textsuperscript{th} Nikolas L{\"o}w}
\IEEEauthorblockA{\textit{Mannheim Institute for Intelligent Systems in Medicine} \\
Heidelberg University\\
Mannheim, Germany \\
nikolas.loew@medma.uni-heidelberg.de}
}
\maketitle

\begin{abstract}
Coded Aperture Imaging (CAI) has been proposed as an alternative collimation technique in nuclear imaging. To maximize spatial resolution small pinholes in the coded aperture mask are required. However, a high-resolution detector is needed to correctly sample the point spread function (PSF) to keep the Nyquist-Shannon sampling theorem satisfied. The disadvantage of smaller pixels, though, is the resulting higher Poisson noise.
Thus, the aim of this paper was to investigate if sufficiently accurate CAI reconstruction is achievable with a detector which undersamples the PSF.
With the Monte Carlo simulation framework TOPAS a test image with multiple spheres of different diameter was simulated based on the setup of an experimental gamma camera from previous work. Additionally, measured phantom data were acquired.
The captured detector images were converted to low-resolution images of different pixel sizes according to the super-resolution factor $k$. 
Multiple analytical reconstruction methods and a Machine Learning approach were compared based on the contrast-to-noise ratio (CNR).
We show, that all reconstruction methods are able to reconstruct both the test image and the measured phantom data for $k \leq 7$. With a synthetic high-resolution PSF and upsampling the simulated low-resolution detector image by bilinear interpolation the CNR can be kept approximately constant.
Results of this simulation study and additional validation on measured phantom data indicate that an undersampling detector can be combined with small aperture holes. However, further experiments need to be conducted. 
\end{abstract}

\begin{IEEEkeywords}
coded aperture imaging, gamma imaging, image reconstruction, super-resolution, nuclear medicine.
\end{IEEEkeywords}

\section{Introduction}

\begin{figure}[bp]
    \centering
    \includegraphics[width=0.9\columnwidth]{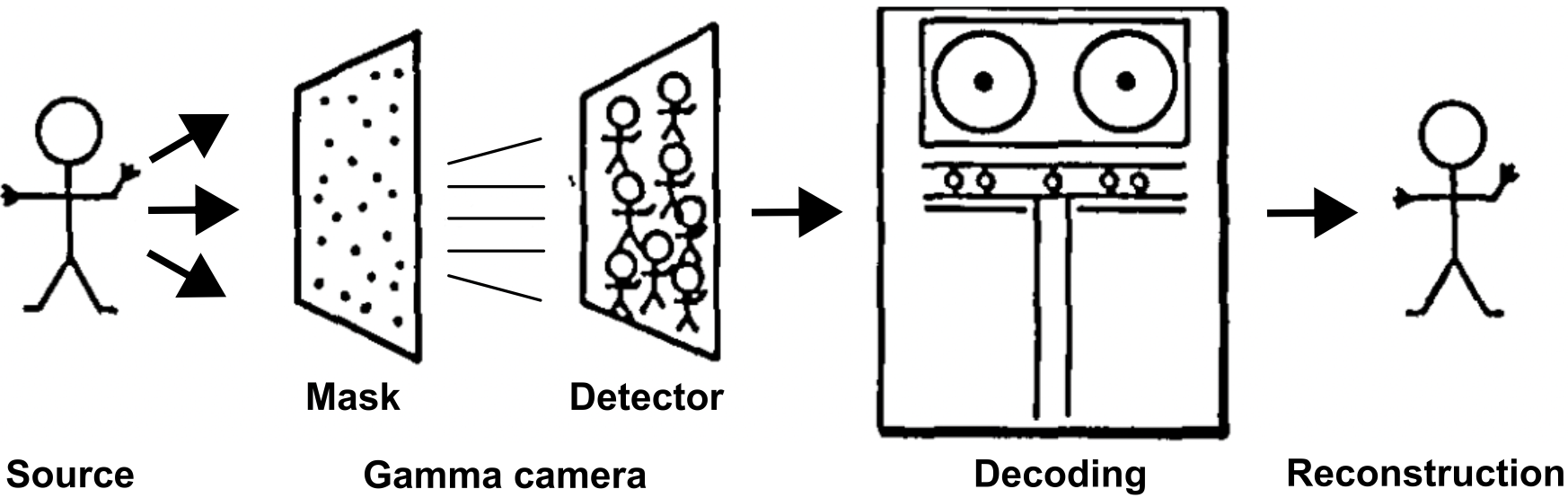}
    \caption{The basic principle of planar Coded Aperture Imaging: A mask with pinholes projects the source image onto the detector, where multiple overlapping projections emerge. Decoding or image reconstruction is necessary to obtain the original source image. Figure modified from~\cite{Accorsi2001a}.}
    \label{fig:my_cai_principle}
\end{figure}

Accurate localization and visualization of radioactive sources is an essential task in nuclear medicine~\cite{Peterson2011}, high-energy astrophysics~\cite{Braga2020} and in monitoring of nuclear waste~\cite{Amgarou2016, Sun2020}. 
Recently, small handheld gamma cameras for localizing sentinel lymph nodes in breast cancer patients are under investigation~\cite{Tsuchimochi2013, Kogler2020, Russo2020, Kaissas2020}.
Due to the high-energy photons involved refractive lenses cannot be used for producing an image of the scene and instead parallel or pinhole collimators are employed to capture the necessary spatial information~\cite{Peterson2011}. 
However, the size of the opening is usually subject to a balanced trade-off between the number of captured photons (photon efficiency) and the spatial resolution as the former increases and the latter decreases with the size of the pinhole. 
A high photon efficiency is desired since the guiding principle in the medical domain is to reduce the exposed radiation to As Low As Reasonable Achievable (ALARA). Thus, photon flux is limited and achieving high quantum yield is of major importance.\\
To improve the mentioned trade-off Coded Aperture Imaging (CAI) has been introduced~\cite{Ables1968, dicke1968scatter}: A mask between object and detector consisting of a radiopaque material with pinholes encodes the directional information of incoming gamma rays. As Figure\,\ref{fig:my_cai_principle} shows, each pinhole in the mask generates a projection of the source image on the detector resulting in a multitude of overlapping projections. Therefore, image reconstruction (also referred to as decoding) becomes necessary. 
If the distance between source and collimator is large and the extension in depth is small relative to the distance, CAI can be considered as an image-to-image mapping and is denoted as planar Coded Aperture Imaging~\cite{Accorsi2001a}.
In this paper, the captured detector image is denoted as $p(x, y)$, the original source image and its reconstruction as $f(x, y)$ and $\hat{f}(x, y)$, respectively. 

\begin{figure}[bpt]
    \centering
    \includegraphics[width=0.9\columnwidth, trim=20 20 10 10,clip]{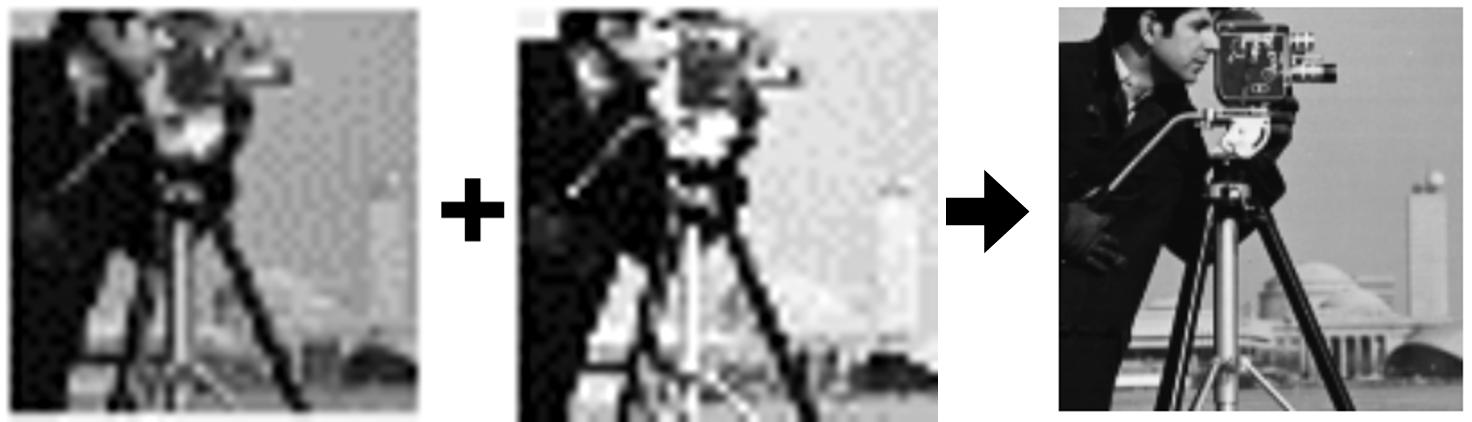}
    \caption{Super-resolution is referred to the process to combine multiple low quality images to reconstruct an image of the underlying high-resolution scene. Figure modified from~\cite{Marcia2008}.}
    \label{fig:super_resolution_principle}
\end{figure}

The term super-resolution refers to the process of combining several "low resolution, noisy, slightly shifted observations"~\cite{Marcia2008} to reconstruct an image of the underlying high resolution scene, as Figure~\ref{fig:super_resolution_principle} illustrates. 

Because the spatial resolution in CAI is mainly influenced by the mask's pinhole diameter~\cite{Kulow2020}, increasing the MURA rank and thus the amount of pinholes while reducing its diameter would increase the spatial resolution. So far, the pinhole diameter has been chosen such that the utilized detector can properly sample the resulting PSF~\cite{Russo2020, Kaissas2020}. \\
To the best of the authors' knowledge, no research group has investigated the combination of small pinholes and a low-resolution detector. 
Therefore, the investigated hypothesis is as follows: Existing CAI reconstruction methods are capable of reconstructing point sources from an undersampling detector, and thus achieving super-resolution, at reasonable quality even though the detector cannot resolve the higher spatial resolution of the aperture. This is due to the shifted but overlapping projections caused by the coded aperture.
\section{Methods}

\subsection{Simulating a coded aperture test image}
For simulating a test image the Monte Carlo simulation toolkit TOPAS~\cite{Perl2012}, a wrapper library around Geant4~\cite{ALLISON2016186}, is deployed. Unlike ray-casting simulations, TOPAS accounts for photon-mass interactions like scattering and mask penetration, and is therefore considered to be the gold standard in gamma imaging~\cite{Rozhkov2020}.\\
The geometrical components and dimensions were simulated according to the experimental gamma camera from Rozhkov et al.~\cite{Rozhkov2020}. The main characteristics can be summarized as follows: A 2$\times$2 mosaicked, 1\,mm thick Tungsten not-two-holes-touching (NTHT) MURA mask of rank 31 with pinholes of 0.34\,mm in diameter (denoted as $d$) was placed 42\,mm ($a$) in front of a 2\,mm thick 256$\times$256 pixelated CdTe semiconductor detector coupled to a Timepix\textsuperscript{\textcopyright} readout circuit. The detector has a side length of 14.1\,mm and hence a single pixel size $s= 0.0551$\,mm. The virtual object plane is 172\,mm ($b$) in front of the mask plane, resulting in a field of view (FoV) of 57.75$\times$57.75\,mm. \\
The test image consists of three spherical sources with diameters $d_1$, $d_2$ and $d_3$ of 1, 2 and 3\,mm distributed within the FoV as Figure~\ref{fig:example_imgs} shows. $10^9$ gamma photons with a photon energy of 140.5\,keV (corresponding to the photon peak of $^{99\text{m}}\text{Tc}$ the most commonly used radiotracer in nuclear medicine~\cite{Peterson2011}) were distributed to the three sources according to their area. Every photon hitting the detector was collected and stored in a so called \textit{phase space file}.
In addition to the coded aperture a single pinhole collimator with the same diameter was simulated to serve as a reference for the reconstructed images. The captured pinhole image was smoothed by Gaussian blurring with a $\sigma$ of 2\,pixels. \\
The ground truth image was generated from the geometrical model and remains binary: 1 for where a source is located and 0 everywhere else. 

\subsection{Measured data from the experimental gamma-camera}
\begin{figure}[bpt]
    \centering
    \includegraphics[width=0.7\columnwidth, trim=0 40 0 40,clip]{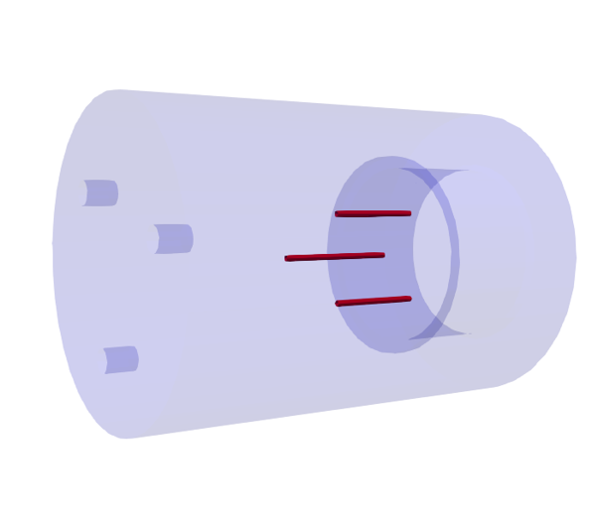}
    \caption{The utilized phantom with its three tubes (red) filled with  $^{99\text{m}}\text{Tc}$ of which 120 images where captured.}
    \label{fig:sr_phantom}
\end{figure}

Captured images from an experimental gamma-camera also used in previous work~\cite{Rozhkov2020, Meißner2023} was used to validate the effect of super-resolution on real-world measurement data. The camera set-up was the same as for the simulation of the test image. The phantom has the basic form of a cylinder with a height of 80\,mm and 50\,mm in diameter, where tubes along the vertical axis were filled with $^{99\text{m}}\text{Tc}$. 
These three tubes have a diameter of 1.1\,mm, and two of them are 15\,mm long while the central one is 20\,mm long. The total activity at the beginning of the measurements was 83\,MBq. A depiction of the geometric computer model can be seen in Figure~\ref{fig:sr_phantom}. The phantom was exposed to the gamma-camera for 2\,min and afterwards rotated by 3\degree. This way, a total of 120 images were captured. Outlier replacement as described in~\cite{Meißner2023} was applied afterwards.

\subsection{Generating low-resolution detector images}
\begin{figure}[tbp]
    \centering
    \includegraphics[width=1.0\columnwidth, trim=0 0 0 0,clip]{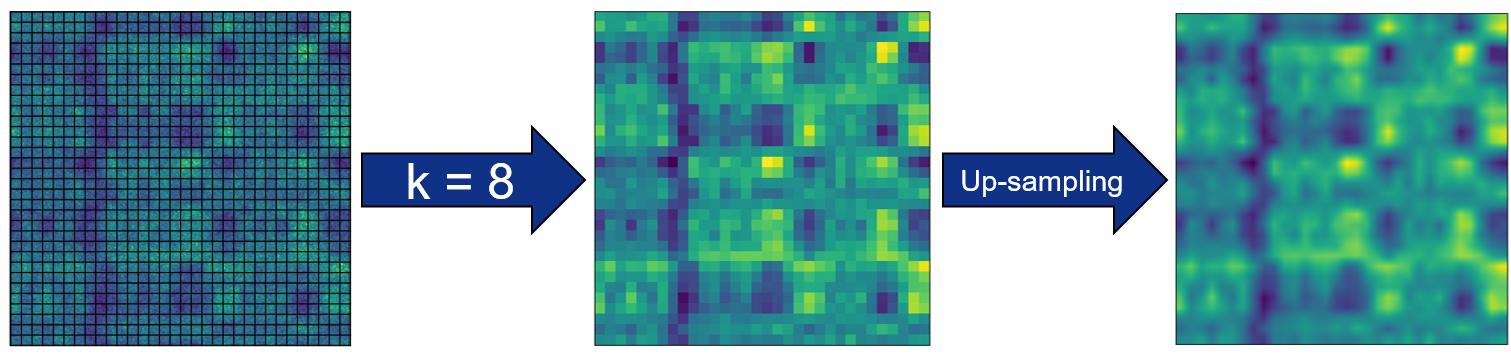}
    \caption{Pixels of the high-resolution detector image from the TOPAS simulation are accumulated (here with $k=8$ into 32$\times$32 pixels) to form the low-resolution detector image. Afterwards this image is upsampled by bilinear interpolation to the high-resolution of 256$\times$256 pixels.}
    \label{fig:down_up_sampling_k8}
\end{figure}

To analyze the effect of different pixel sizes low-resolution images of different resolutions were produced as follows: The captured photons from the phase space file and from the measured phantom data respectively were binned into images of different low resolution. The actual detector served as reference with a resolution of 256$\times$256 pixels, which corresponds to the resolution of the final reconstructed image. Therefore, the super-resolution factor $k$ is introduced. It means that $k\times k$ high-resolution pixels are reconstructed from a single low-resolution pixel. Note that the absolute detector size remains the same: 14.1$\times$14.1\,mm. The single pixel size $s$ changes proportional to $k$: $s=k\cdot14.1\,$mm$/256$. Finally, all low-resolution images were upsampled by bilinear interpolation to 256$\times$256 pixels in order to fit the synthetic high-resolution PSF and to fit into the CED-IN respectively. 
The process of generating low-resolution images is shown in Figure~\ref{fig:down_up_sampling_k8} for $k=8$.

\subsection{Analytical reconstruction methods}

\begin{figure}[tbp]
    \centering
    \includegraphics[width=1.0\columnwidth, trim=0 0 0 0,clip]{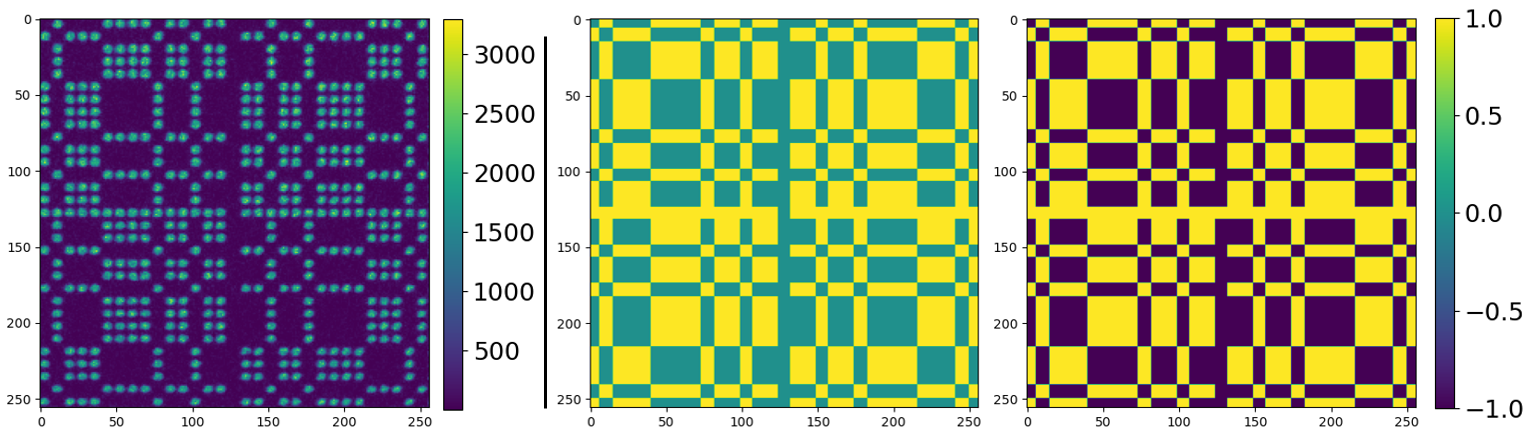}
    \caption{Left: The measured point spread function (PSF) of the experimental gamma camera, where the pixel intensity represents photon counts. Center: The two-holes-touching (THT) version of the used rank 31 MURA mask $h(x, y)$ fundamental for MURA Decoding and MLEM. Right: The respective decoding pattern $g(x, y)$ for MURA Decoding. Note the additional positive square at the center of $g(x, y)$~\cite{Cieslak2016}. Both patterns were resized to 256$\times$256 pixels by nearest neighbor interpolation to maintain the original shape.}
    \label{fig:mask_patterns3}
\end{figure}

Three different methods for super-resolution reconstruction are analyzed and compared in this paper: MURA decoding~\cite{Fenimore1978}, a convolutional Maximum Likelihood Expectation Maximization algorithm (MLEM)~\cite{Mu2006} and a convolutional encoder-decoder network (CED) from previous work~\cite{Meißner2023}. \\
\indent MURA Decoding is the most commonly used reconstruction method. It consists of a single circular convolution of the detector image $p(x, y)$ with the decoding pattern $g(x, y)$:

\begin{equation}
    \hat{f}(x,y) = p(x,y) \circledast g(x,y) 
    \label{eq:MURA_decoding}
\end{equation}

\noindent with "$\circledast$" denoting the circular convolution operator. All circular operations in this paper are carried out by periodically padding the second operand to twice its size, i.e. to 512$\times$512 pixels, and cropping the result to its central 256$\times$256 pixels.\\
The decoding pattern $g(x, y)$ is based on $h(x, y)$ and its definition can be found in~\cite{Cieslak2016}: It is equivalent to changing all $0$ to -$1$ and adding a positive pixel to the center of the PSF~\cite{Cieslak2016}. \\
The MLEM algorithm works in iterations and is derived from a random Poisson process. It consists of a combination of forward and backward projections, where ten iterations were deployed in this paper:

\begin{equation}
\label{eq:mlem}
    \hat{f}^{k+1}(x,y) = \hat{f}^{k}(x,y) \odot \left[\frac{p(x,y)}{ \hat{f}^{k}(x,y) \circledast h(x,y)} \otimes h(x,y)\right]
\end{equation}

\noindent where "$\odot$" denotes point-wise multiplication and "$\otimes$" is circular cross-correlation. \\
Instead of the real measured PSF with round pinhole projections, the two-holes-touching (THT) version of the PSF without gaps between neighboring pinholes is used for reconstruction, since it suppresses periodical noise~\cite{Meißner2023}. Both the THT-PSF and its corresponding decoding pattern are of rectangular structure and a square of 8 bright pixels represents the position and bounding box of each projected pinhole. They also define the reconstructed resolution of 256$\times$256 pixels. Figure~\ref{fig:mask_patterns3} depicts the measured PSF, the THT-PSF and the decoding pattern. \\
Additionally, MURA Decoding with low-resolution THT-PSF was implemented, where the synthetic high-resolution THT-PSF was not used, but the down-sampled THT-PSFs emulating a low-resolution detector. Since the reconstructions come in low-resolution, the reconstructed images were upsampled to 256$\times$256 pixels by bilinear interpolation.

\subsection{Reconstruction by Machine Learning}
\begin{figure}[tbp]
    \centering
    \includegraphics[width=0.9\columnwidth, trim=0 0 0 0, clip]{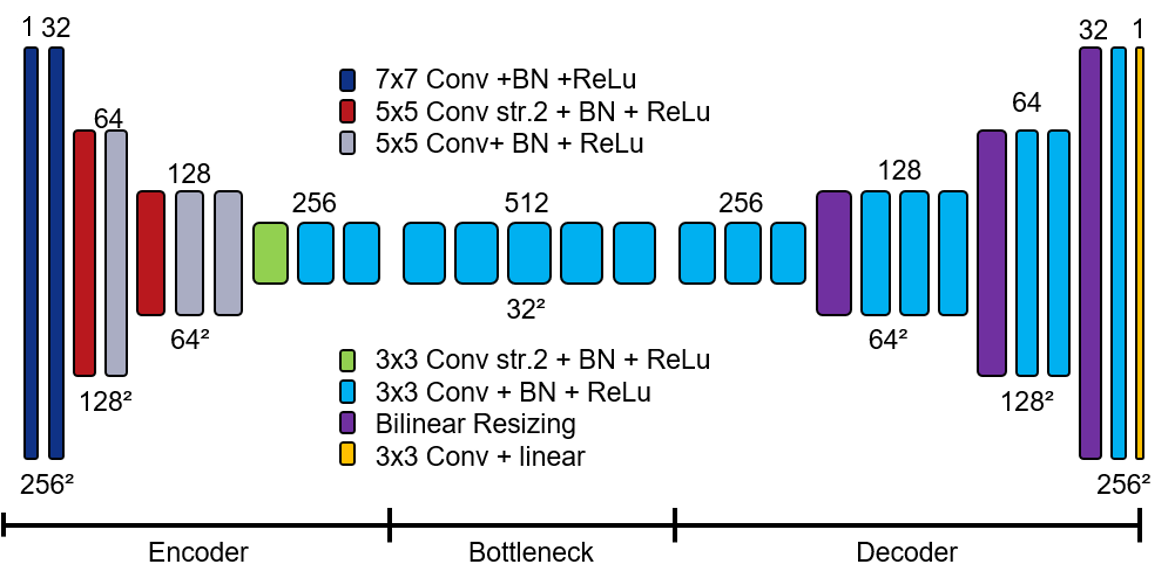}
    \caption{The convolutional encoder-decoder network architecture deployed in this paper. The top row represents the number of filters per layer and the bottom row the feature map size in pixels.}
    \label{fig:ced_architecture}
\end{figure}

A Convolutional Encoder-Decoder (CED) is a widely used form of Convolutional Neural Networks (CNN). A CED consists of trainable parameters that transform an input into an output image. However, this transformation is not derived from a mathematical description but by providing a sufficient amount of paired training images. 
First experiments were conducted on the application of CNNs to CAI reconstruction, but its validation exclusively relied on simulated and low-resolution images~\cite{Zhang2019} or were only visually compared on few images~\cite{Kulow2020}.
While the two analytical reconstruction methods solely rely on the PSF of the gamma camera, which acts as a linear approximation of the imaging system, the CED-IN is in theory capable of more complex mappings~\cite{Goodfellow2016}. Recent advances in Machine Learning in the field of image reconstruction \cite{Haggstrom2019, Belthangady2019, Zhu2018} underline the potential of CEDs for CAI reconstruction.
The CED used in this paper is denoted as CED-IN because it was trained with a convolutional simulation based on natural photographs from the ImageNet database~\cite{Russakovsky2015}. Its architecture is presented in Figure~\ref{fig:ced_architecture} and for more information on the training process and data simulation the reader is referred to~\cite{Meißner2023}. 

After reconstructing all images, the contrast-to-noise ratio (CNR) is calculated based on the reconstructed and the ground truth image. The binary ground truth image enables a separation of the reconstruction into the signal part $S$ and background part $B$. The following definition of CNR is employed \cite{Zhang2020}:

\begin{equation}
\label{eq:cnr}
    \text{CNR} = \frac{\left|\bar{S} - \bar{B}\right|}{\sigma_B}\,, 
\end{equation}

\noindent where $\bar{S}$ denotes the mean intensity of the signal, $\bar{B}$ the mean intensity and $\sigma_B$ the standard deviation of the background.

\begin{figure*}[htb]
\centering
\begin{subfigure}[b]{0.85\textwidth}
   \includegraphics[width=1\linewidth]{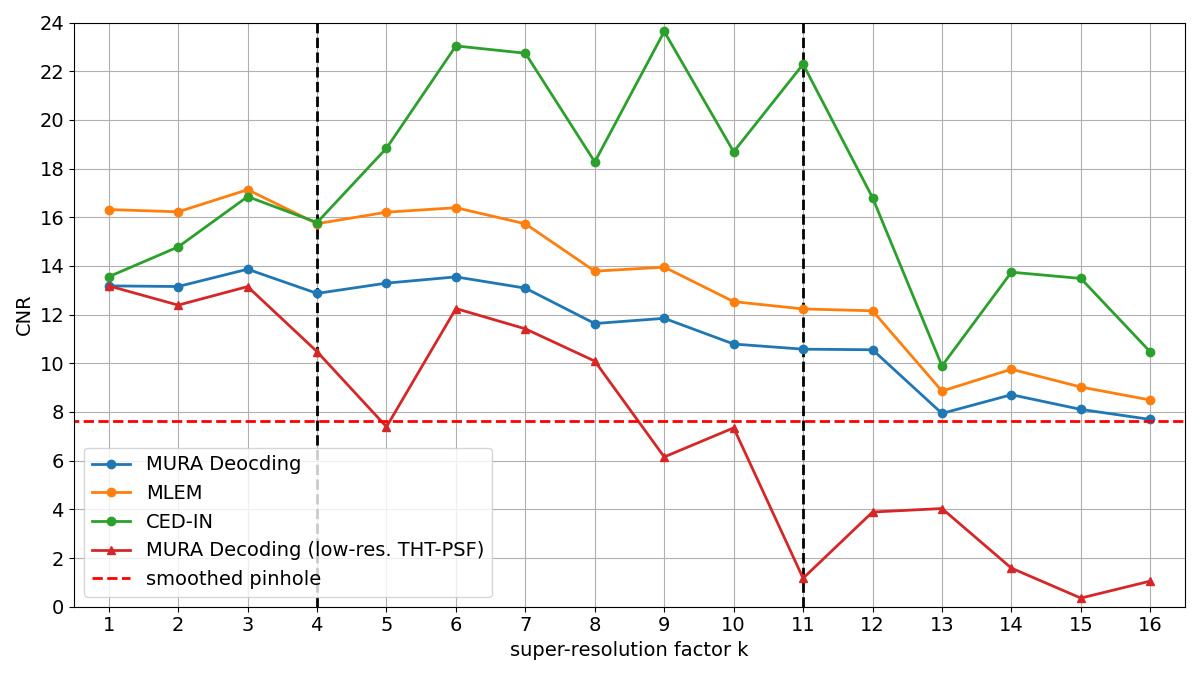}
    \caption{Simulated test image: The black dotted vertical lines mark the critical super-resolution factors $\tilde{k}_{\text{THT-PSF}}=4$ and $\tilde{k}_1=11$.The red dotted line represents the CNR of the smoothed image captured by a pinhole collimator and serves as reference.}
    \label{fig:cnr_vs_k_with_lr_psf}
\end{subfigure}
\begin{subfigure}[b]{0.85\textwidth}
   \includegraphics[width=1\linewidth]{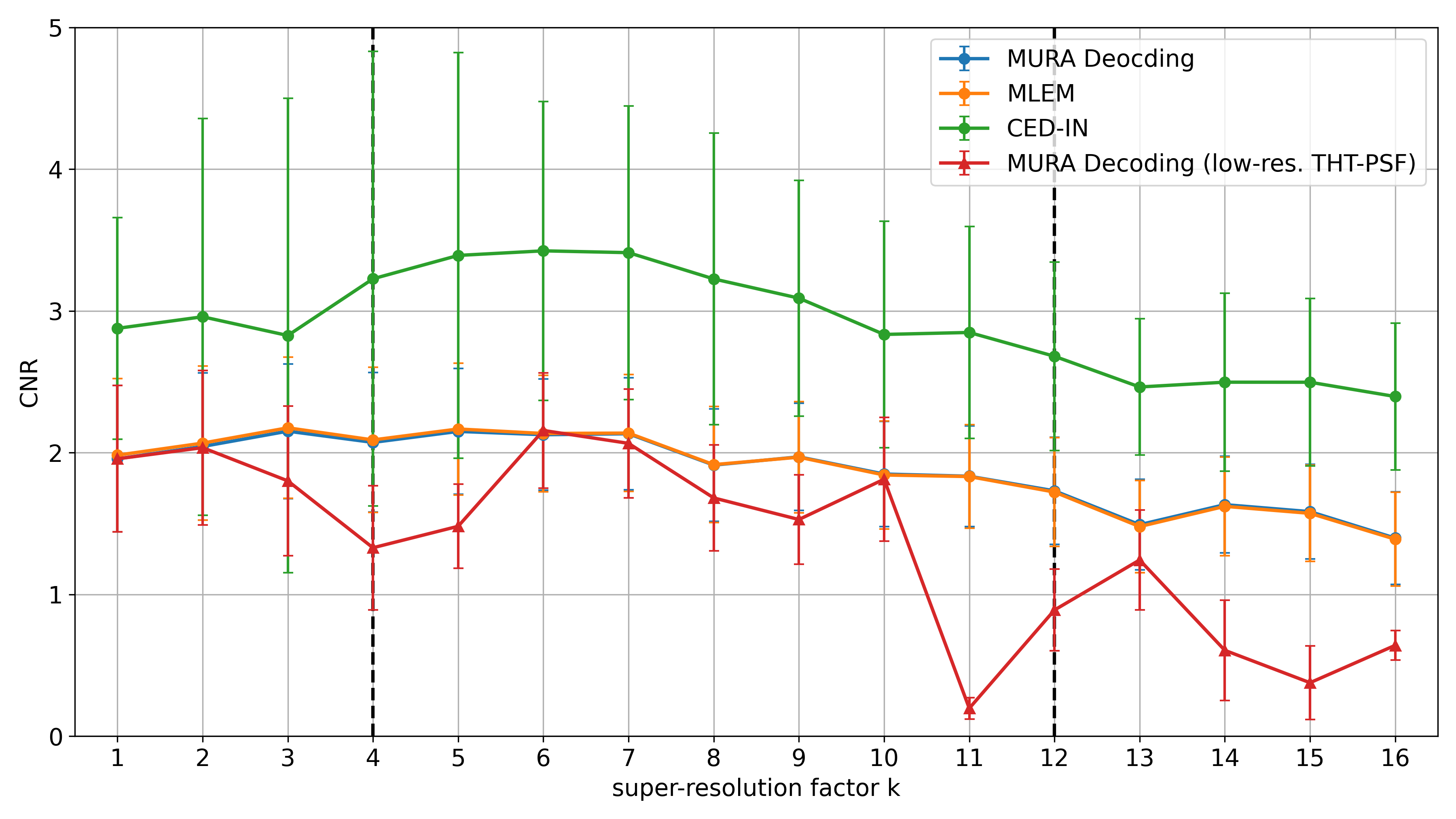}
    \caption{Measured phantom data: Error bars represent the standard deviation at each data point. The vertical dotted lines denote the critical super-resolution factors $\tilde{k}_{\text{THT-PSF}}=4$ and $\tilde{k}_{\text{measurement}}=12$.}
    \label{fig:phantom_cnrs}
\end{subfigure}
\caption{The contrast-to-noise ratio (CNR) for reconstructions of different reconstruction methods depending on the super-resolution factor $k$. }
\end{figure*}

\begin{figure*}[htb]
\centering
\begin{subfigure}[b]{0.95\textwidth}
\includegraphics[width=0.85\textwidth, trim=25 10 15 10, clip]{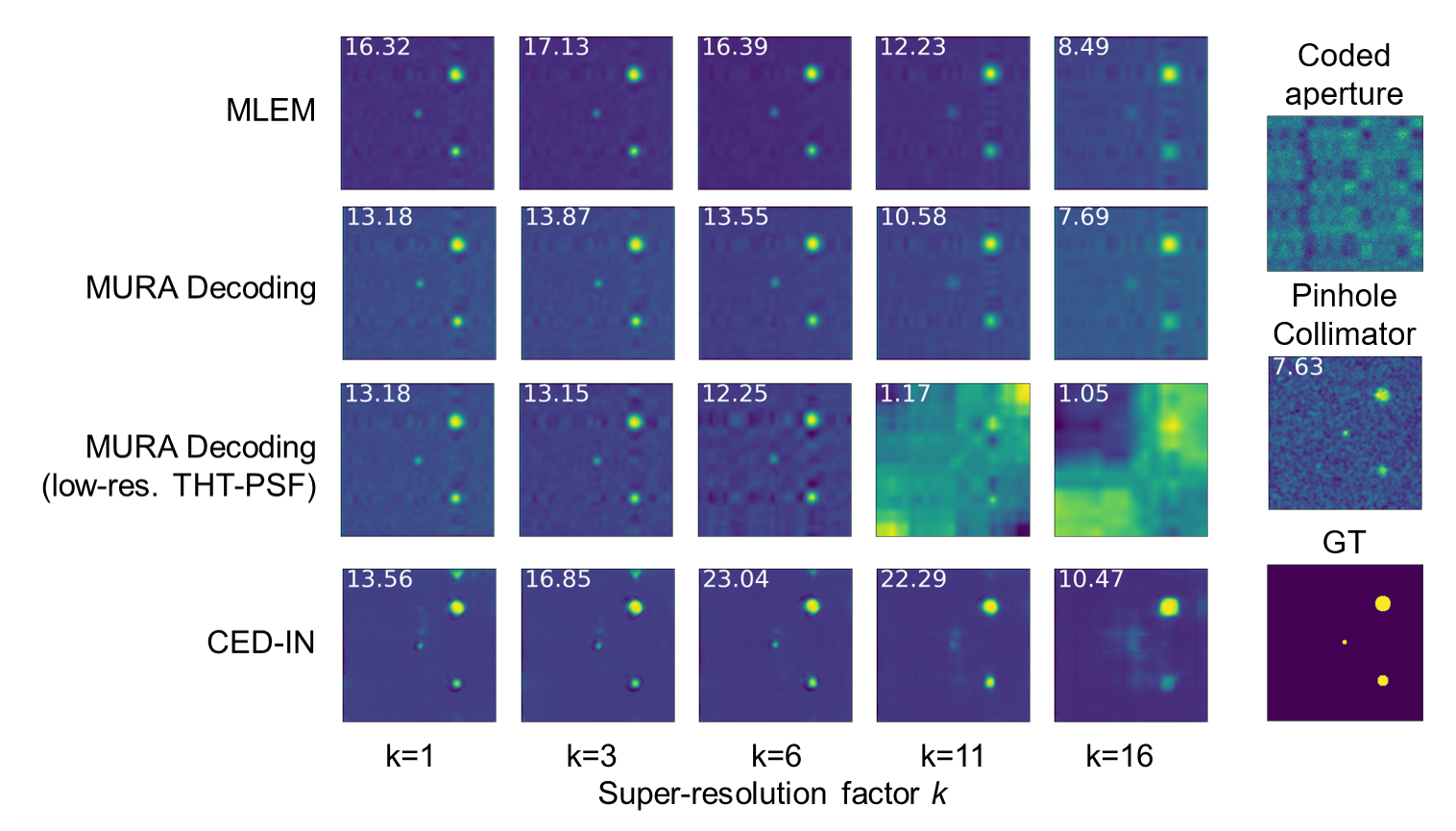}
    \caption{Simulated test image: For reference, the coded aperture simulation in 256$\times$256 pixels, the smoothed pinhole collimator simulation in the same resolution and the ground truth is shown on the right-hand side.}
    \label{fig:example_imgs}
\end{subfigure}
\begin{subfigure}[b]{0.95\textwidth}
       \includegraphics[width=0.85\textwidth, trim=25 10 15 10, clip]{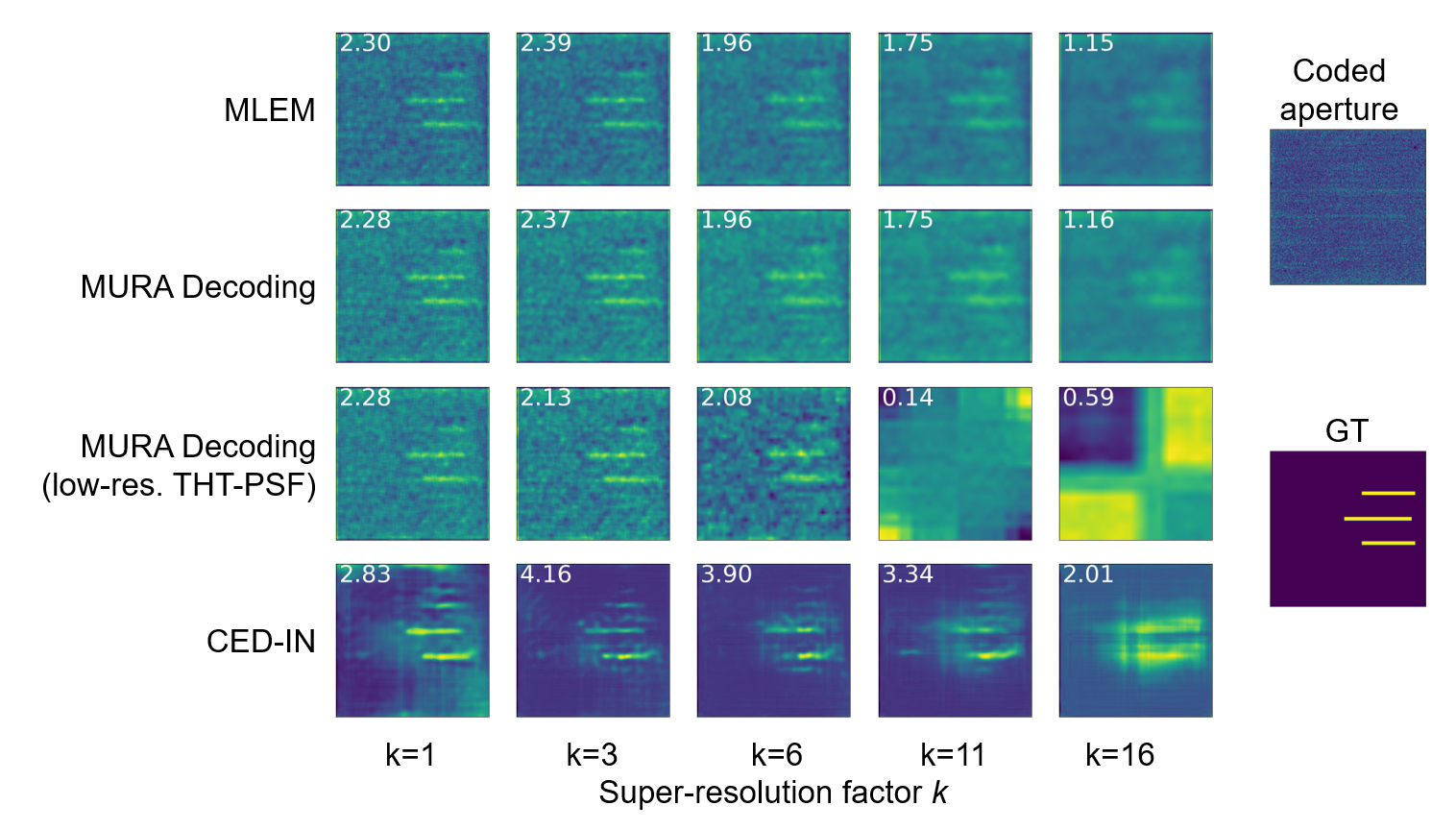}
        \caption{Measured phantom data: On the right-hand side the coded aperture detector and the ground truth (GT) images are shown.}
    \label{fig:phantom_ex_images}
\end{subfigure}
\caption{Exemplary reconstructions of the analyzed reconstruction methods at different super-resolution factors $k$. The CNR is printed in the top left corner of each image. }
\end{figure*}

\subsection{Nyquist-Shannon sampling theorem}
The Nyquist-Shannon sampling theorem states that the sampling frequency of a pixelated representation must be larger than twice the maximum frequency of the periodic image~\cite{beyerer2015machine}. Thus, when the smallest occurring structure is sampled by two pixels or less, an image is not represented unambiguously which leads to aliasing and hence signal degradation~\cite{beyerer2015machine}.
Since the aforementioned analytical reconstruction methods MURA Decoding and MLEM consist of one or more convolutions of two discretized signals, a reconstruction without aliasing artefacts is only possible when both images were sampled by enough pixels. Thus, critical super-resolution factors $\tilde{k}$ were determined both for the coded aperture test image and the THT-PSF $h(x, y)$. The smallest point source of the test image is 1\,mm wide and therefore much larger than the pinhole diameter: $d_1 \gg d$. Hence, the smallest structure on the detector caused by the small point source can be approximated by $t = d_1 \cdot m = 1.244$\,mm. For $h(x, y)$ the smallest structure $t$ is 8 pixels wide, i.e. $t=8 \cdot s$ (Figure~\ref{fig:mask_patterns3}). For the given gamma camera with its magnification factor $m = (1+a/b) = 1.244$, the smallest depicted structure $t$ and the single pixel side length of $s = 0.0551$\,mm $\tilde{k}$ can be defined as follows:

\begin{equation}
\label{eq:k_crit}
    \tilde{k} = \left\lfloor \frac{1}{2} \cdot \frac{t}{s} \right\rfloor
\end{equation}

\noindent where $\left\lfloor \cdot \right\rfloor$ denotes rounding off to the next smallest integer value.\\

\section{Results}
\subsection{Critical super-resolution factors}
The following critical super-resolution factors $\tilde{k}$ were obtained from the Nyquist-Shannon sampling theorem: The THT-PSF $h(x, y)$ must be sampled by at least 64$\times$64 pixels leading to the following critical super-resolution factor: $\tilde{k}_{\text{THT-PSF}} = 4$. This means that the synthetic high-resolution THT-PSF in this paper is 4-times oversampled. However, the test image with the 1\,mm point source results in a higher critical super-resolution factor of $\tilde{k}_{1} = \left\lfloor 11.29 \right\rfloor = 11$.
The tubes of the phantom captured by the experimental gamma-camera have a diameter of 1.1\,mm, that are magnified to approximately 1.37\,mm and thus 24.83 pixels. This results in a critical super-resolution factor for the measured data of $\tilde{k}_{\text{measurement}} = \left\lfloor 12.42 \right\rfloor = 12$.\\

\subsection{Results on the test image}
Figure~\ref{fig:cnr_vs_k_with_lr_psf} shows the CNR of the four reconstruction methods over the super-resolution factor $k$. The red dotted line at $\text{CNR}=7.63$ denotes the smoothed image captured with a pinhole collimator where no reconstruction was required. It is depicted central on the right-hand side together with the ground truth image and the coded aperture test image in Figure~\ref{fig:example_imgs}. The left-hand side shows exemplary reconstructions for $k=1, 3, 6, 11$ and $16$. \\
Clearly visible, the CNRs of the reconstruction methods with the synthetic high-resolution THT-PSF increase until $k=3$ and steadily decline afterwards. The CED-IN is an exception, where the CNR increases further until falling below its baseline at $k=13$. 
For all reconstruction methods using the synthetic high-resolution THT-PSF the smallest point source starts to disappear for $k > 11$ and is hardly visible for $k=16$. \\
MURA Decoding with the respective low-resolution THT-PSF does not exceed its baseline CNR and falls beneath the pinhole reference at $k=5$ and again for all $k>9$. The reconstructions for $k \geq 11$ fail entirely and resemble no similarity to the ground truth anymore.

\subsection{Results on the measurement data}
Analogously to the test image Figure~\ref{fig:phantom_cnrs} shows the CNR for the presented reconstruction method at different super-resolution factors $k$. However, because 120 reconstructions were analyzed the marker represents the average CNR and additional error bars represent the standard deviation for each reconstruction method and super-resolution factor $k$.\\
In general, compared to the test image, lower CNRs can be observed. Similarly, all methods except for the MURA Decoding with low-resolution THT-PSF, slightly rise for small $k$ and then fall after approximately $k = 7$. This behavior can also be seen in the exemplary images in Figure~\ref{fig:phantom_ex_images}. The maximum median CNR is reached by the CED-IN with 3.42 at super-resolution factor $k=6$. 
Visually, a higher background noise is present compared to the test image and the three line sources are prominent until they start to disappear for larger $k$.

\section{Discussion}
\subsection{Simulated test image}
For the given setup the Nyquist-Shannon sampling theorem states, that for super-resolution factors above $k=4$ the PSF is not sufficiently sampled anymore. If the THT-PSF is undersampled it loses its characteristic to properly function for CAI reconstruction. The simulation study of this paper shows this behavior where CNRs of MURA Decoding with low-resolution THT-PSF drop notably for $k \geq \tilde{k}_{\text{THT-PSF}}$ and the reconstruction of $k=11$ show major artefacts rendering the three sources unrecognizably.
But, when upsampling the low-resolution detector image to a high-resolution of 256$\times$256 pixels and using a synthetic high-resolution THT-PSF for reconstructions, the CNRs drop slower, as MURA Decoding and MLEM demonstrate. For $k \leq 6$ the CNRs even stay approximately constant. \\
Additionally, the reconstructed images are visually closer to the expected output. For $k > \tilde{k}_1$, as predicted by the Nyquist-Shannon sampling theorem, the smallest point source cannot be reconstructed properly and starts to dissolve. \\
In this simulation study super-resolution with small $k$ and upsampling by bilinear interpolation even had a positive impact on the CNR and the reconstructions look smoother. 

\subsection{Measured phantom data}
Similar behavior was observed for the measured phantom data. For all reconstruction methods, the CNR does not decrease until approximately $k = 7$, even though undersampling of the THT-PSF starts at $k = 4$. However, the gain in CNR for the measured phantom data is far less for small $k$ compared to the simulated test image. \\
Interestingly, the CED-IN, even though not trained on processing upsampled low-resolution images, performs better than all other reconstruction methods for super-resolution factors of $k \geq 4$. For the measured phantom data it is the best reconstruction method for all $k$. This indicates that the CED-IN generalized from the training domain of natural photographs to discrete sources on a dark background, even when the low-resolution input image was upsampled by bilinear interpolation. Especially the background is reconstructed more uniform compared to all other methods. \\ 
This implies that the CED-IN was taught to compress the inputted detector image into a robust representation of the image, suppressing noise and dead pixels, like in the top right corner of the coded aperture image in Figure~\ref{fig:phantom_ex_images}.\\
The test image generated by Monte Carlo simulation shows in theory that super-resolution in CAI is possible and the simulated low-resolution detector images based on phantom data captured by an experimental gamma-camera strengthen this hypothesis. 
However, the question remains as to how a real low-resolution detector would affect the reconstruction. Erroneous pixels on a low-resolution detector will have a higher impact since it is supposed to capture a larger fraction of the coded aperture pattern. Another point not investigated in this paper are other types of gamma sources. Especially extended sources are known to cause problems in CAI~\cite{Mu2016}. 

\section{Conclusion}
The conducted simulation study indicates that super-resolution reconstruction for planar CAI is possible even if the detector is not capable of sampling the PSF with a sufficient amount of pixels. Instead, a synthetic high-resolution THT-PSF is combined with upsampling the captured low-resolution detector image by bilinear interpolation. This way, established reconstruction methods were able to reconstruct the simulated test image. 
However, for large super-resolution factors, the smallest point source could not be reconstructed as the Nyquist-Shannon sampling theorem predicted. 
Applying the same technique to simulated low-resolution detector images from data of a hot-rod phantom captured with an experimental gamma-camera strengthen these findings. For future research, though, further experiments with a more realistic undersampling detector including erroneous pixels are needed.


\bibliographystyle{ieeetr}
\bibliography{references}

\end{document}